\documentclass[aps,prd,showpacs,showkeys,notitlepage,preprintnumbers,nofootinbib]{revtex4-1}
\usepackage{amsmath,amssymb,amsthm,bm,color,hyperref,blindtext,fullpage}
\usepackage{graphics,subfigure,rotating,etoolbox,amsbsy,graphicx}
\usepackage[utf8]{inputenc}
\usepackage[toc,page]{appendix}
\usepackage[dvipsnames]{xcolor}
\newcommand{\nocontentsline}[3]{}
\newcommand{\tocless}[2]{\bgroup\let\addcontentsline=\nocontentsline#1{#2}\egroup}

\begin{document} 

\title{Energy of a point-like neutron
in an external electromagnetic field}

\author{Jesus Saenz}
\email{jmsaenzv@nmsu.edu}
\affiliation{Institute of Engineering and Technology,
Universidad Aut\'onoma de Ciudad Ju\'arez, 32310, Ciudad Ju\'arez, Mexico}
\affiliation{Department of Physics, New Mexico State University,
Las Cruces, NM 88003, USA}
\author{Michael Engelhardt}
\email{engel@nmsu.edu}
\affiliation{Department of Physics, New Mexico State University,
Las Cruces, NM 88003, USA}
\author{Roman H\"ollwieser}
\email{hoellwieser@uni-wuppertal.de}
\affiliation{Department of Physics, University of Wuppertal,
42119 Wuppertal, Germany}
\affiliation{Department of Physics, New Mexico State University,
Las Cruces, NM 88003, USA}

\begin{abstract}
A point-like neutron in an external electromagnetic field experiences a
shift in energy that mimicks the effect of an actual structural deformation
of an extended neutron, i.e., a proper polarizability. In order to be able
to differentiate between the former and the latter, a Foldy-Wouthuysen
transformation is constructed which yields the energy shift of a point-like
neutron quadratic in the external field in a derivative expansion,
generalizing a long-known result for the dipole electric polarizability
due to Foldy. The ten leading Foldy contributions to the energy are
determined for a zero-momentum neutron. In addition, eliminating the
momentum operator in favor of the velocity operator, analogous results
are derived for a zero-velocity neutron. In this case, operator ordering
ambiguities are encountered that permit only a determination of eight of
the ten Foldy terms.
\end{abstract}

% \pacs{11.15.Ha, 12.38.Gc}


\maketitle

\section{Introduction}
\label{introsec}
%%%%%%%%%%%%%%%%%%%%%
Electromagnetic fields polarize nucleons by coupling to the electric
charges of their quark constituents. The degree to which nucleons are
susceptible to these polarization effects constitutes a basic question
about nucleon structure. For sufficiently weak fields, such effects are
quantified through polarizabilities, which characterize the linear
response of the nucleon to the electromagnetic field; in terms of an
effective Hamiltonian, polarizabilities are coefficients of terms
quadratic in the electromagnetic field.

In addition, the fields can be classified according to their space-time
variation, starting with the most basic case of constant electric and
magnetic fields that induce the so-called dipole polarizabilities;
generalizing to space-time dependent fields, the effective Hamiltonian
can then be organized into a derivative expansion. The leading
polarizability-related terms of the effective Hamiltonian in the
expansion in space-time derivatives of the electromagnetic field
read\footnote{At variance with \cite{babusci}, the present work employs
Gaussian units.} \cite{babusci}
\begin{eqnarray}
H_{eff} &=& -\frac{1}{2} \left[ \alpha_{E} E^2 \ +\beta_{M} B^2 \
+\gamma_{E1} \sigma \cdot (E \times \dot{E} ) \
+\gamma_{M1} \sigma \cdot (B \times \dot{B} ) \
-2\gamma_{E2} E_{ij} \sigma_i B_j \
+2\gamma_{M2} B_{ij} \sigma_i E_j \right. \nonumber \\
& & \left. \ \ \ \ \ \ +\alpha_{E\nu } \dot{E}^2 \
+\beta_{M\nu } \dot{B}^2 \
+\frac{1}{6} \alpha_{E2}E_{ij}^2 \
+\frac{1}{6} \beta_{M2}B_{ij}^2
+ \ldots \right]
\label{heff}
\end{eqnarray}
with the quadrupole strengths of the electric and magnetic fields
\begin{equation}
E_{ij}=\frac{1}{2}(\nabla _i E_j+\nabla_j E_i) \ ,
\ \ \ \ \ \ B_{ij}=\frac{1}{2}(\nabla _i B_j+\nabla_j B_i) \ .
\end{equation}
Here, $\alpha_{E} $ and $\beta_{M} $ are the aforementioned dipole
electric and magnetic polarizabilities; further, $\gamma_{E1} $,
$\gamma_{M1} $, $\gamma_{E2} $, and $\gamma_{M2}$ are the spin
polarizabilities, $\alpha_{E\nu } $ and $\beta_{M\nu }$ are the
dispersion polarizabilities, and $\alpha_{E2} $ and $\beta_{M2}$
are the quadrupole polarizabilities.

It should be noted that casting the response of nucleons to external fields
in the local, truncated form (\ref{heff}) constitutes a significant
assumption. In general, the energy of the nucleon in the presence of external
fields is a nonlocal functional incorporating information about space-time
correlations between the nucleon wave function and the external field.
Adopting the form (\ref{heff}) implies that the nucleon wave function
is sufficiently localized such that the nucleon's energy only
depends on the local values of the electromagnetic field and its first
derivatives; in (\ref{heff}), those quantities are Taylor coefficients
of an expansion of the external field around the position of the nucleon,
and as such are not themselves functions of space and time anymore,
for a given nucleon position. On the other hand, the assumption of a
localized nucleon state quantum mechanically clashes with the notion
of studying the nucleon at rest, which is generically associated with
a spatially extended wave function. The consequent limitations on the
applicability of the form (\ref{heff}) will constitute an important
aspect of the present study; extended comments follow further below.

As laid out in detail in \cite{babusci}, the ten polarizabilities
in (\ref{heff})
can be connected to the amplitudes for nucleon Compton scattering in
the low-energy limit. Continued progress in the phenomenological
extraction of electromagnetic polarizabilities, which more recently has
begun to encompass not only the dipole polarizabilities, but also
spin polarizabilities, has been reported in
\cite{holst1,pasq1,grie1,schuma,pasq5,pasq6,grie2,grie3,holst2,myers,martel1,gryn1,gryn2,hagel,grie4,pasq2,pasq3,pasq4,miskimen,martel2,paudyal,melen}.
Sensitivity to orders beyond the ones displayed in (\ref{heff}) was
considered in \cite{pasq6,gryn2}. On the other hand, efforts have been
undertaken to evaluate polarizabilities in Lattice QCD, by computing
hadron mass shifts in the presence of external electromagnetic fields
\cite{woloshyn,wilepap,leepap,shintani,elpol,polprog,detlat1,detlat2,spinpol,alee,leea,gwgamma,lujan1,freeman,luschev,lujan2,adel1,adel2,adel3,adel4,adel5,adel6}.
Chiral Effective Theory serves to connect such lattice data, which are
obtained at heavier-than-physical pion masses on finite volumes, to the
physical limit \cite{grie1,detcsb,lensky,grie5,adel5}. In the case of magnetic
fields, care must be taken to disentangle the mass shift from the Landau
level structure \cite{tiburzi1,adel1,adel2,adel3,adel4,adel5,adel6}, and effects beyond
linear response may contaminate the analysis \cite{tiburzi2,tiburzi3}.
The subtleties involved in matching the background field calculations
performed in Lattice QCD to the Effective Field Theory description of
scattering amplitudes were examined in detail in \cite{jwlee1,jwlee2}.

The interpretation of hadron mass shifts in terms of polarizabilities is not
bereft of subtlety. Already before the advent of the current understanding
of nucleon structure in terms of underlying quark and gluon degrees of
freedom, it was noted by Foldy \cite{foldy} that even a point-like neutron
in the presence of a constant electric field experiences an energy shift
quadratic in the electric field. The argument can be made quite succinctly:
Writing the Dirac equation for a neutral point particle with an
(entirely anomalous) magnetic moment $\mu $ as
\begin{equation}
\left[ i\gamma^{\mu } \partial_{\mu } -\frac{\mu }{2} \sigma_{\mu \nu }
F^{\mu \nu } -m \right] \psi = 0 \ ,
\end{equation}
the corresponding Dirac Hamiltonian reads (cf.~eq.~(\ref{diracconv}) below
for the Dirac structure conventions employed in this work)
\begin{equation}
H=\alpha \cdot p +i\mu \gamma \cdot E - \mu \, \Pi \cdot B + \beta m
\label{dirach}
\end{equation}
If the external field is purely electric and constant, $[H,p]=0$; then,
the energy of a zero-momentum neutron, $W^{p=0}$, can be extracted by
noting that, for $p=0$, one has $H^2 = m^2 + \mu^{2} E^2 $, and hence
\begin{equation}
W^{p=0} = \sqrt{m^2 + \mu^{2} E^2 } = m + \frac{\mu^{2} E^2 }{2m} + \ldots
\label{foldyshift}
\end{equation}
The energy shift quadratic in the electric field $E$ mimics the
effect of a polarizability, cf.~(\ref{heff}). In effect,
$\alpha_{E}^{Foldy} = -\mu^{2} /m$. However, it is not due to an actual
structural deformation of the neutron; the neutron was treated as a
point particle (at most, one may argue that the anomalous magnetic
moment $\mu $ is chiefly a consequence of the neutron's substructure).
It may therefore be useful to separate this effect from the effect due
to an actual polarization of the neutron \cite{coon1,detlat2}, i.e.,
subtract the term proportional to $E^2 $ on the right-hand side of
(\ref{foldyshift}) from the mass shift of the neutron obtained in
a constant external electric field, in order to extract the dipole
electric polarizability proper. Correspondingly, standard phenomenological
analyses apportion this contribution to the Born, non-structure parts
of the amplitudes describing nucleon Compton scattering \cite{grie2,grie5}.

A constant magnetic field $B$, for which solutions of the Dirac equation
for particles with and without electric charge and with anomalous magnetic
moment are discussed in detail in \cite{pitsch}, cf.~also \cite{strocchi},
does not induce a Foldy-type term analogous to the electric one in
(\ref{foldyshift}). Solutions of the Dirac equation for neutral
particles in more general forms of magnetic field are discussed
in \cite{shish}, without, however, allowing for a direct identification
of Foldy-type coefficients.

Motivated by the advent of experimental data allowing one to extract
spin polarizabilities \cite{martel1,miskimen,martel2,paudyal}, the
purpose of the present work is to expand the treatment of
Foldy-type effects for a neutron from the simple dipole polarizability
cases highlighted above to all ten polarizabilities defined in
eq.~(\ref{heff}). This is achieved by way of constructing an appropriate
Foldy-Wouthuysen transformation \cite{fwtrafo}. An additional aspect
that will be taken into account is the one stressed in \cite{coon1,coon2},
namely, that in the presence of electromagnetic fields, zero momentum
and zero velocity are not synonymous. The Foldy contributions to the
energy of a neutron in both types of states will be considered. In the
case of a zero-velocity neutron, obstructions will be encountered that
ultimately only allow one to determine eight of the ten Foldy-type
coefficients. These obstructions appear to be symptoms of the general
limitation of the definition (\ref{heff}) already noted further above,
namely, that the form (\ref{heff}) implies a localization of the neutron
wave function that clashes with the notion of studying the neutron at
rest, be it in the sense of zero momentum or zero velocity.

Further expanding upon this latter point, in extracting a local result of
the form (\ref{heff}) from an initial description of the neutron in terms
of a Dirac Hamiltonian, cf.~(\ref{dirach}), with external fields depending
on space and time, one must take care when invoking the localization
assumption. At first sight, a neutron state can certainly become localized
despite its momentum being limited to negligible magnitudes as long as it
is sufficiently heavy. As one takes the local limit, corrections due to
the residual extent of the neutron wave function will vanish as an
inverse power of the neutron mass. However, also the Foldy terms at
issue here vanish as an inverse power of the neutron mass. Therefore,
it is necessary to compare the behavior of the two effects carefully as
the local limit is taken. This is explored in section \ref{perturbsec} and,
indeed, the two effects are of the same order. Therefore, the aforementioned
corrections due to the residual extent of the neutron wave function
must be taken into account in any comprehensive analysis of a concrete
physical setting, such as, say, a Lattice QCD calculation of the neutron
energy (along with, of course, the many other systematic effects arising
in such a setting). Nevertheless, it should be emphasized that these
corrections depend on further details of the environment in which the
neutron is placed, such as boundary conditions, and thus do not
constitute intrinsic electromagnetic properties of the neutron on
the same footing as the Foldy contributions.

As indicated by this preliminary discussion, the emphasis of the present
study lies as much on ascertaining the boundaries of a description in terms
of a local effective Hamiltonian of the form (\ref{heff}) as it does on
extracting concrete results for the Foldy contributions associated with
the polarizabilities in (\ref{heff}) to the extent possible. These
limitations will become apparent in more than one aspect, and to exhibit
them is as much a goal of this investigation as is the determination of
those Foldy-type effects that are accessible in a such a framework.

%%%%%%%%%%%%%%%%%%%%%

\section{Foldy-Wouthuysen transformation}

\subsection{General form of the transformation}
%%%%%%%%%%%%%%%%%%%%%
The Foldy-Wouthuysen transformation \cite{fwtrafo,sil} serves to
decouple the dynamics of the particle and antiparticle components
of a Dirac spinor, at least to a given order in an expansion scheme.
Expansion in the inverse particle mass, $1/m$, yields the relativistic
corrections to the non-relativistic Hamiltonian; here, the expansion
parameters will instead be the external electromagnetic field strengths
along with their derivatives. To achieve this expansion, the following
treatment will largely follow the scheme laid out in \cite{sil}. Consider a
Hamiltonian of the form
\begin{equation}
H=\beta m + {\cal E} + {\cal O},
\label{htemplate}
\end{equation}
with even and odd operators characterized by
$\beta {\cal E}={\cal E}\beta$, $\beta {\cal O}=-{\cal O}\beta$, inducing
an equation of motion
\begin{equation}
\left( -i\frac{\partial }{\partial t} + H \right) \psi =0.
\end{equation}
The goal is to transform this (for the purpose of mitigating the effects
of ${\cal O} $) as
\begin{equation}
U^{\dagger } \left( -i\frac{\partial }{\partial t} +
H \right) U \psi^{\prime } =0,
\end{equation}
where $\psi^{\prime } = U^{\dagger } \psi $. The new Hamiltonian can now
be extracted by observing that
\begin{equation}
\left[ -i\frac{\partial }{\partial t}
+U^{\dagger } \left( -i\frac{\partial U }{\partial t} \right)
+U^{\dagger } H U \right]
\psi^{\prime } =0,
\end{equation}
i.e., one has the new Hamiltonian
\begin{equation}
H^{\prime } = U^{\dagger } \left( -i\frac{\partial U }{\partial t} \right)
+U^{\dagger } (\beta m + {\cal O} )U
+U^{\dagger } {\cal E} U.
\end{equation}
In \cite{sil}, the following transformation is constructed,
\begin{equation}
U=\frac{\epsilon + m -\beta {\cal O} }{\sqrt{2\epsilon (\epsilon +m)} } \ ,
\label{trafo}
\end{equation}
where $\epsilon = \sqrt{m^2 + {\cal O}^{2} } $. Note that one is largely
free in the ordering of the different parts of this operator; merely
the relative ordering of the $\beta $ and $ {\cal O} $ factors in the
numerator matters. Otherwise, the different parts commute. Furthermore,
$\epsilon $ is positive definite, so there are no problems defining
square roots and inverses. One can easily check unitarity,
$U^{\dagger } U =1$, and also
\begin{equation}
U^{\dagger } (\beta m + {\cal O} )U = \beta \epsilon = \beta m +
\beta (\epsilon -m).
\end{equation}
Hence, one has succeeded in eliminating the odd term ${\cal O} $ in the
Hamiltonian in favor of the even term $\beta (\epsilon -m)$; the
new Hamiltonian now reads
\begin{equation}
H^{\prime } = \beta m + \beta (\epsilon -m)
+U^{\dagger } \left( -i\frac{\partial U }{\partial t} \right)
+U^{\dagger } {\cal E} U.
\label{newham}
\end{equation}
However, the other two terms in $H^{\prime } $ may reintroduce new odd
terms, i.e., in general, this transformation is not exact. Nonetheless,
if ${\cal O} $ and ${\cal E} $ are in some sense small, i.e., if one
is content with a power expansion, the induced new odd terms may be of
higher order, and it will be sufficient to iterate the transformation
a finite number of times, until the remaining odd terms are of
sufficiently high order to be dropped. For present purposes, it will
be necessary to keep only terms of up to second order in ${\cal E } $,
or second order in $\partial / \partial t$, or altogether fourth order
in the objects ${\cal E } $, ${\cal O } $, $\partial / \partial t$.
This specification will be justified below as more concrete expressions
become available from which one can read off the required order.
Also the second derivative $\ddot{\cal O } $ can be dropped. Note that,
here and in the following, the dot denotes the {\em partial} derivative
$\partial / \partial t$. Expanding
\begin{eqnarray}
\epsilon &=& m\left( 1+\frac{1}{2} \frac{ {\cal O}^{2} }{m^2 }
-\frac{1}{8} \frac{ {\cal O}^{4} }{m^4 } + \ldots \right) \\
U &=& 1 - \frac{1}{2m} \beta {\cal O} - \frac{1}{8m^2 } {\cal O}^{2}
+\frac{3}{16m^3 } \beta {\cal O}^{3} + \frac{11}{128m^4 } {\cal O}^{4}
+ \ldots \\
U^{\dagger } \left( -i\frac{\partial U }{\partial t} \right) &=&
\frac{i}{2m} \beta \dot{\cal O} +\frac{i}{8m^2 }
[\dot{\cal O} , {\cal O} ] - \frac{i}{16m^3 } \beta
\left( 3 {\cal O }^{2} \dot{\cal O} + 2 {\cal O } \dot{\cal O} {\cal O }
+ 3 \dot{\cal O} {\cal O }^{2} \right)
+ \ldots \\
U^{\dagger } {\cal E} U &=& {\cal E} +
\frac{1}{2m} \beta [ {\cal O} , {\cal E} ]
+\frac{1}{8m^2 } \left[ [ {\cal O} , {\cal E} ] , {\cal O} \right]
+\frac{1}{16m^3 } \beta \left( 3 [ {\cal E} , {\cal O}^{3} ]
+[ {\cal O} , {\cal O} {\cal E} {\cal O} ] \right)
+ \ldots
\end{eqnarray}
and classifying the terms with respect to their even/odd character, the
new Hamiltonian is
\begin{equation}
H^{\prime } = \beta m + {\cal E}^{\prime } +{\cal O}^{\prime }
\end{equation}
with
\begin{eqnarray}
{\cal E}^{\prime } &=& \frac{1}{2m} \beta {\cal O}^{2}
-\frac{1}{8m^3 } \beta {\cal O}^{4} + \frac{i}{8m^2 }
[\dot{\cal O} , {\cal O} ] + {\cal E}
+\frac{1}{8m^2 } \left[ [ {\cal O} , {\cal E} ] , {\cal O} \right] \\
{\cal O}^{\prime } &=& \frac{i}{2m} \beta \dot{\cal O}
- \frac{i}{16m^3 } \beta
\left( 3 {\cal O }^{2} \dot{\cal O} + 2 {\cal O } \dot{\cal O} {\cal O }
+ 3 \dot{\cal O} {\cal O }^{2} \right)
+\frac{1}{2m} \beta [ {\cal O} , {\cal E} ]
+\frac{1}{16m^3 } \beta \left( 3 [ {\cal E} , {\cal O}^{3} ]
+[ {\cal O} , {\cal O} {\cal E} {\cal O} ] \right)
\end{eqnarray}
The odd term ${\cal O}^{\prime } $ now starts at one order higher in
${\cal E } $, ${\cal O } $, $\partial / \partial t$ than the original
${\cal O} $. This is not yet sufficient to preclude additional contributions
to the even term up to the desired order upon further iteration. Repeating
the procedure, one has the new Hamiltonian
\begin{equation}
H^{\prime \prime } = \beta m + {\cal E}^{\prime \prime }
+ {\cal O}^{\prime \prime }
\end{equation}
with
\begin{eqnarray}
{\cal E}^{\prime \prime } &=& \frac{1}{2m} \beta {\cal O}^{\prime 2}
-\frac{1}{8m^3 } \beta {\cal O}^{\prime 4} + \frac{i}{8m^2 }
[\dot{\cal O}^{\prime } , {\cal O}^{\prime } ] + {\cal E}^{\prime }
+\frac{1}{8m^2 } \left[ [ {\cal O}^{\prime } , {\cal E}^{\prime } ] ,
{\cal O}^{\prime } \right] \nonumber \\
&=& \frac{1}{2m} \beta {\cal O}^{2}
-\frac{1}{8m^3 } \beta {\cal O}^{4} + \frac{i}{8m^2 }
[\dot{\cal O} , {\cal O} ] + {\cal E}
+\frac{1}{8m^2 } \left[ [ {\cal O} , {\cal E} ] , {\cal O} \right]
\nonumber \\
& & + \frac{1}{8m^3 } \beta \dot{\cal O}^{2}
-\frac{1}{8m^3 } \beta [ {\cal O} , {\cal E} ]^{2}
-\frac{i}{8m^3 } \beta \dot{\cal O} [ {\cal O} , {\cal E} ]
-\frac{i}{8m^3 } \beta [ {\cal O} , {\cal E} ] \dot{\cal O} \\
{\cal O}^{\prime \prime } &=& \frac{i}{2m} \beta \dot{\cal O}^{\prime }
- \frac{i}{16m^3 } \beta
\left( 3 {\cal O }^{\prime 2} \dot{\cal O}^{\prime }
+ 2 {\cal O }^{\prime } \dot{\cal O}^{\prime } {\cal O }^{\prime }
+ 3 \dot{\cal O}^{\prime } {\cal O }^{\prime 2} \right)
+\frac{1}{2m} \beta [ {\cal O}^{\prime } , {\cal E}^{\prime } ]
+\frac{1}{16m^3 } \beta \left( 3 [ {\cal E}^{\prime } , {\cal O}^{\prime 3} ]
+[ {\cal O}^{\prime } ,
{\cal O}^{\prime } {\cal E}^{\prime } {\cal O}^{\prime } ] \right)
\nonumber \\
&=& \frac{i}{2m^2 } [ \dot{\cal O} , {\cal E} ]
+\frac{i}{4m^2 } [ {\cal O} , \dot{\cal E} ]
-\frac{i}{8m^3 } \beta
( {\cal O}^{2} \dot{\cal O} + \dot{\cal O} {\cal O}^{2} )
-\frac{1}{4m^2 } \left[ {\cal E} , [ {\cal O} , {\cal E} ] \right]
-\frac{1}{8m^3 } \beta \left( {\cal O}^{2} [ {\cal O} , {\cal E} ]
+ [ {\cal O} , {\cal E} ] {\cal O}^{2} \right)
\end{eqnarray}
The odd term ${\cal O}^{\prime \prime } $ now starts at third order in
${\cal E } $, ${\cal O } $, $\partial / \partial t$. This means that
subsequent iterations will only contribute new terms to the even part
that are of too high order to be retained, while successively increasing
the order of the odd part until it can be completely dropped. Thus, to
the desired order, the final Foldy-Wouthuysen Hamiltonian has been
obtained,
\begin{eqnarray}
H_{FW} \ \ = \ \ \beta m + {\cal E}^{\prime \prime } &=&
\beta m + \frac{1}{2m} \beta {\cal O}^{2}
-\frac{1}{8m^3 } \beta {\cal O}^{4} + \frac{i}{8m^2 }
[\dot{\cal O} , {\cal O} ] + {\cal E}
+\frac{1}{8m^2 } \left[ [ {\cal O} , {\cal E} ] , {\cal O} \right]
\nonumber \\
& & + \frac{1}{8m^3 } \beta \dot{\cal O}^{2}
-\frac{1}{8m^3 } \beta [ {\cal O} , {\cal E} ]^{2}
-\frac{i}{8m^3 } \beta \left( \dot{\cal O} [ {\cal O} , {\cal E} ]
+ [ {\cal O} , {\cal E} ] \dot{\cal O} \right)
\label{hfw}
\end{eqnarray}

%%%%%%%%%%%%%%%%%%%%%

\subsection{Evaluation in terms of background fields and momenta}
\label{evalsec}
%%%%%%%%%%%%%%%%%%%%%
The Dirac Hamiltonian for the neutron with anomalous magnetic moment
$\mu $ is of the form (\ref{htemplate}), with, cf.~(\ref{dirach}),
\begin{equation}
{\cal O} = \alpha \cdot p +i\mu \gamma \cdot E \ ,
\ \ \ \ \ \ \ \ {\cal E} = -\mu \Pi \cdot B
\end{equation}
Here, the position representation is adopted, where $p=-i\nabla $, and the
Dirac representation is used, in which
\begin{equation}
\beta = \left( \begin{array}{cc} 1 & 0 \\ 0 & -1 \end{array} \right)
\ \ \ \ \ \ \ \
\alpha^{i} = \left(
\begin{array}{cc} 0 & \sigma^{i} \\ \sigma^{i} & 0 \end{array} \right)
\ \ \ \ \ \ \ \
\gamma^{i} = \beta \alpha^{i}
\ \ \ \ \ \ \ \
\Pi^{i} = \beta \sigma^{i}
\ \ \ \ \ \ \ \
\gamma^{5} = \left( \begin{array}{cc} 0 & 1 \\ 1 & 0 \end{array} \right)
\label{diracconv}
\end{equation}
with $\sigma^{i} $ denoting the Pauli matrices.
The goal is to present the Foldy-Wouthuysen Hamiltonian $H_{FW} $ in a
form in which:
\begin{itemize}
\item
All momentum operators $p$ have been commuted through to the right.
\item
The Dirac structures have been simplified to a manifestly block-diagonal
form, upon which $H_{FW} $ can be restricted to the upper components,
leaving at most Pauli matrix structures.
\item
Terms higher than quadratic in the external fields $E$, $B$ have been dropped.
\item
Terms containing higher than first derivatives of external fields have
been dropped.
\item
Terms of higher than altogether fourth order in the objects
$E$, $B$, $\partial /\partial t$, $\nabla $, $p$ have been dropped (where
$\partial /\partial t$, $\nabla $ are always acting only on a specific
external field, whereas $p$ stands as an operator on its own).
\item
At most one power of $p$ is kept in terms already quadratic in the external
fields $E$, $B$.
\item
At most three powers of $p$ are kept in terms linear in the external 
fields $E$, $B$.
\end{itemize}
Again, the reasoning leading to this specification will become fully
apparent below as the treatment unfolds; of course, the third
and fourth points are already clear from the stated objectives of the
calculation, i.e., expansion to second order in the external fields,
up to first derivatives of those fields. Note that this specification
justifies the truncation imposed in the previous section, as stated
after eq.~(\ref{newham}): Since ${\cal E} $ is proportional to $B$,
only up to second order in ${\cal E} $ is required; since only one derivative
each of at most two external fields is allowed, only up to second order in
$\partial /\partial t$ is required; and since both ${\cal E} $ and
${\cal O} $ each supply one order in the objects $E$, $B$, $\nabla $, $p$,
only up to fourth order in ${\cal E } $, ${\cal O } $, $\partial / \partial t$
is required.

Treating the terms appearing in (\ref{hfw}) in turn, one has
\begin{equation}
{\cal O}^{2} = -\mu \nabla \cdot E -i\mu \sigma \cdot (\nabla \times E)
+\mu^{2} E^2 -2\mu \sigma \cdot ( E\times p) + p^2
\label{o2}
\end{equation}
where, having restricted to the upper components, one can set $\beta =1$.
Also, here, and in the following, the derivative $\nabla $ only acts on
the field immediately to its right, whereas the momentum operator
$p=-i\nabla $ acts on all objects to its right. The term ${\cal O}^{4} $,
which is the most complex one appearing in $H_{FW} $, can be obtained by
squaring (\ref{o2}),
\begin{eqnarray}
{\cal O}^{4} &=& -\mu^{2} \frac{1}{2} (\nabla_{i} E_j +\nabla_{j} E_i )^2
-\mu^{2} (\nabla_{i} E_j -\nabla_{j} E_i )^2 + \mu^{2} (\nabla \cdot E)^2
+2\mu^{2} i\sigma \cdot (\nabla \times E) (\nabla \cdot E) \nonumber \\
& & +4\mu^{2} \sigma_{j} E_i \nabla_{j} E_l \epsilon_{ilm} p_m
-4\mu^{2} (\sigma \cdot E) (\nabla \times E) \cdot p
-12\mu^{2} i E_i \nabla_{j} E_i p_j +4\mu^{2} i E_i \nabla_{i} E_j p_j
\nonumber \\
& & +4\mu^{2} (\nabla \cdot E) \sigma \cdot (E\times p)
+4\mu^{2} i (\nabla \cdot E) E \cdot p
+4\mu i \sigma_{k} \epsilon_{kij} \nabla_{l} E_i p_l p_j
-2\mu i \sigma \cdot (\nabla \times E) p^2 \nonumber \\
& & -2\mu (\nabla \cdot E) p^2
-4\mu \sigma \cdot (E \times p) p^2 + p^4
\end{eqnarray}
Arriving at this result requires only standard, if lengthy, Pauli matrix
and $\epsilon $-symbol algebra, apart, perhaps, from the not immediately
apparent identity $(\nabla \times E)_{j} (\nabla_{i} E_j -\nabla_{j} E_i )=0$.
One furthermore has
\begin{eqnarray}
[ \dot{\cal O} , {\cal O} ] &=&
-2\mu^{2} i \sigma \cdot (E\times \dot{E} ) + 2\mu i \dot{E} \cdot p \\
\dot{\cal O}^{2} &=& \mu^{2} \dot{E}^{2}
\end{eqnarray}
Turning to the terms containing ${\cal E} $, apart from the original
even operator
\begin{equation}
{\cal E} = -\mu \sigma \cdot B
\end{equation}
all other terms involve the intermediate odd operator
\begin{equation}
[ {\cal O} , {\cal E} ] = \mu \beta \gamma^{5} \sigma \cdot (\nabla \times B)
+2\mu \beta \gamma^{5} B\cdot p +2\mu^{2} i \gamma^{5} E\cdot B
\end{equation}
where Maxwell's equations for the external fields have been used to
drop $\nabla \cdot B =0$. From this, one obtains
\begin{eqnarray}
\left[ [ {\cal O} , {\cal E} ] , {\cal O} \right] &=&
-\mu^{2} E_j \sigma_{i} \left( 3(\nabla_{i} B_j - \nabla_{j} B_i )
+ (\nabla_{i} B_j + \nabla_{j} B_i ) \right)
-2\mu^{2} B_i \sigma_{j} (\nabla_{i} E_j + \nabla_{j} E_i ) \nonumber \\
& & +2\mu (\nabla \times B) \cdot p -2\mu i \sigma_{j} \nabla_{j} B_i p_i
+4\mu \sigma_{j} B_i p_j p_i \\
\left[ {\cal O} , {\cal E} \right]^{2}
&=& -\mu^{2} (\nabla \times B)^{2}
+4\mu^{2} i B_i \nabla_{i} B_j p_j -4\mu^{2} \sigma \cdot (\nabla \times B)
B\cdot p \\
\dot{\cal O} [ {\cal O} , {\cal E} ] + [ {\cal O} , {\cal E} ] \dot{\cal O}
&=& -2\mu^{2} i (\nabla \times B) \cdot \dot{E}
-4\mu^{2} i (\sigma \cdot \dot{E} ) B\cdot p
\label{odotoe}
\end{eqnarray}
Inserting (\ref{o2})-(\ref{odotoe}) into (\ref{hfw}) yields the
Foldy-Wouthuysen Hamiltonian expressed in terms of background fields
and momenta.
For further use, it is convenient to gather the aggregate
expression\footnote{The difference in sign between the term
$-\frac{1}{2m} \mu \nabla \cdot E$ in (\ref{hfwall}) and the
corresponding term in \cite{coon1} should be noted. This difference
can be traced back to a corresponding difference in sign of the
electric field term in the original Dirac Hamiltonian (\ref{dirach})
compared to the one used in \cite{coon1}. To compare expressions
between here and \cite{coon1}, one must change the sign of the
magnetic moment $\mu $; Foldy contributions to polarizabilities,
which are proportional to $\mu^{2} $, are not affected.},
\begin{eqnarray}
H_{FW} &=& m -\mu \sigma \cdot B + \frac{1}{2m} \left(
-\mu \nabla \cdot E -i\mu \sigma \cdot (\nabla \times E)
+\mu^{2} E^2 -2\mu \sigma \cdot ( E\times p) + p^2
\right) +\frac{1}{8m^3 } \mu^{2} \dot{E}^{2} \nonumber \\
& & -\frac{1}{8m^3 } \left(
-\mu^{2} \frac{1}{2} (\nabla_{i} E_j +\nabla_{j} E_i )^2
-\mu^{2} (\nabla_{i} E_j -\nabla_{j} E_i )^2 + \mu^{2} (\nabla \cdot E)^2
+2\mu^{2} i\sigma \cdot (\nabla \times E) (\nabla \cdot E) \right.
\nonumber \\
& & \ \ \ \ \ \ \ \ \ \ \ \
+4\mu^{2} \sigma_{j} E_i \nabla_{j} E_l \epsilon_{ilm} p_m
-4\mu^{2} (\sigma \cdot E) (\nabla \times E) \cdot p
-12\mu^{2} i E_i \nabla_{j} E_i p_j +4\mu^{2} i E_i \nabla_{i} E_j p_j
\nonumber \\
& & \ \ \ \ \ \ \ \ \ \ \ \
+4\mu^{2} (\nabla \cdot E) \sigma \cdot (E\times p)
+4\mu^{2} i (\nabla \cdot E) E \cdot p
+4\mu i \sigma_{k} \epsilon_{kij} \nabla_{l} E_i p_l p_j
-2\mu i \sigma \cdot (\nabla \times E) p^2 \nonumber \\
& & \ \ \ \ \ \ \ \ \ \ \ \
\left. -2\mu (\nabla \cdot E) p^2
-4\mu \sigma \cdot (E \times p) p^2 + p^4 \right) \nonumber \\
& & +\frac{1}{8m^2 } \left(
2\mu^{2} \sigma \cdot (E\times \dot{E} ) - 2\mu \dot{E} \cdot p \right)
-\frac{1}{8m^3 } \left(
2\mu^{2} (\nabla \times B) \cdot \dot{E}
+4\mu^{2} (\sigma \cdot \dot{E} ) B\cdot p \right) \nonumber \\
& & +\frac{1}{8m^2 } \left(
-\mu^{2} E_j \sigma_{i} \left( 3(\nabla_{i} B_j - \nabla_{j} B_i )
+ (\nabla_{i} B_j + \nabla_{j} B_i ) \right)
-2\mu^{2} B_i \sigma_{j} (\nabla_{i} E_j + \nabla_{j} E_i ) \right.
\nonumber \\
& & \ \ \ \ \ \ \ \ \ \ 
\left. +2\mu (\nabla \times B) \cdot p
-2\mu i \sigma_{j} \nabla_{j} B_i p_i
+4\mu \sigma_{j} B_i p_j p_i \right) \nonumber \\
& & -\frac{1}{8m^3 } \left(
-\mu^{2} (\nabla \times B)^{2}
+4\mu^{2} i B_i \nabla_{i} B_j p_j -4\mu^{2} \sigma \cdot (\nabla \times B)
B\cdot p \right)
\label{hfwall}
\end{eqnarray}

%%%%%%%%%%%%%%%%%%%%%

\section{Energy of zero-momentum neutron states}
%%%%%%%%%%%%%%%%%%%%%
\subsection{Zero momentum and local limit}
In the Foldy-Wouthuysen Hamiltonian (\ref{hfwall}), the
electric and magnetic fields are still functions of space and time. As
already indicated in the preliminary discussion at the end of section
\ref{introsec}, one must take care in reducing this to a result of the form
(\ref{heff}) for the energy of a zero-momentum neutron, in which the
fields and their derivatives are taken to be local constants at the
position of a localized neutron state. For any finite neutron mass, the
neutron wave function must feature a residual extent to allow the neutron
momentum to be bounded to negligible magnitudes. This residual extent
probes the external field in the neighborhood of the neutron, leading
to corrections to the neutron energy compared to the one obtained by
simply replacing all fields in (\ref{hfwall}) directly by their local
values. It should be emphasized that these corrections depend on further
details of the environment in which one places the neutron, which influence
its spatial wave function. They are therefore not purely intrinsic
properties of the neutron that could be discussed in full generality;
instead, they require a model for the environment. Thus, in order to
exemplify these effects, in the detailed evaluation in section
\ref{perturbsec}, the neutron will be placed in a box with periodic
boundary conditions, with the goal of ultimately expressing the neutron
energy in terms of the values of the external fields at the center of
the box. This choice of model is relevant, e.g., for the analysis of
Lattice QCD calculations of neutron energies; furthermore, it allows
one to maintain the notion of an exact zero-momentum state, i.e., one
can simply set $p=0$ when applying (\ref{hfwall}) to such a state.
The $p=0$ state moreover is symmetric about the center of the box,
such that, placing the origin at the center of the box, one has
$\langle p=0 | x_i | p=0 \rangle =0$ in any direction $i$; this reduces
the number of corrections that must be considered. In fact, only two terms
in (\ref{hfwall}) require closer scrutiny to the order being pursued here,
namely, the terms $-\mu \sigma \cdot B$ and $\frac{1}{2m} \mu^{2} E^2 $.
Expanding the spatial dependence of the fields to linear order around the
origin, one has
\begin{eqnarray}
-\mu \sigma \cdot B (x,t) &=& -\mu \sigma \cdot B -
\mu \sigma_{i} (\nabla_{j} B_i ) x_j + \ldots
\label{bxtcorr} \\
\frac{\mu^{2} }{2m} (E(x,t))^2 &=& \frac{\mu^{2} }{2m} \left( E^2
+(\nabla_{j} E_i \nabla_{k} E_i ) x_j x_k \right) + \ldots
\label{extcorr}
\end{eqnarray}
where fields without space-time arguments are evaluated at the origin (the
temporal dependence will be commented upon presently). The space-dependent
terms on the right-hand sides of (\ref{bxtcorr}) and (\ref{extcorr}) spoil
the eigenstate character of the $p=0$ state. Their effects can, however,
be taken into account in perturbation theory around the $p=0$ state. This
will be pursued in section \ref{perturbsec}.

In addition to their spatial dependence, the external fields also are
time-dependent. The time dependence must be treated on a different
footing than the spatial dependence: As far as the latter is concerned,
there is a direct tension between the localization assumption and using
a $p=0$ state, with the uncertainty relation limiting the accuracy to
which both of these specifications can be maintained. Consequently,
taking the spatial dependences in (\ref{bxtcorr}) and (\ref{extcorr})
into account perturbatively cannot be avoided. On the other hand, there
is no analogous direct clash in the temporal direction; if the temporal
dependence of the external fields is sufficiently slow, an adiabatic
treatment becomes applicable, in which transitions to excited states are
absent and the fields can indeed simply be evaluated at the expansion
point $t=0$. In fact, on the contrary, such an adiabatic treatment is
the furthest one can go while remaining consistent with the notion of
a state with a well-defined energy embodied in the initial form
(\ref{heff}) under consideration. Once transitions to excited states
are induced via time-dependent perturbation theory, the form (\ref{heff})
loses its meaning.

In summary, the temporal evolution will be assumed to be adiabatic; the
evaluation of the spatial perturbative corrections is deferred to the
next section. Before taking these corrections into account, one obtains
the neutron energy in a zero-momentum state by simply setting $p=0$ in
the Foldy-Wouthuysen Hamiltonian (\ref{hfwall}),
\begin{eqnarray}
W_{FW}^{p=0} &=& m -\mu \sigma \cdot B
-\frac{\mu }{2m} i\sigma \cdot (\nabla \times E)
-\frac{\mu }{2m} \nabla \cdot E
\nonumber \\
& & +\frac{\mu^{2} }{2m} E^2 +\left[ \frac{\mu^{2} }{4m^2 }
\sigma \cdot (E\times \dot{E} ) -\frac{3\mu^{2} }{8m^2 }
\sigma \cdot (E\times (\nabla \times B)) \right]
\nonumber \\
& & -\frac{\mu^{2} }{8m^2 } \sigma_{i} E_j (\nabla_{i} B_j +\nabla_{j} B_i )
-\frac{\mu^{2} }{4m^2 } \sigma_{i} B_j (\nabla_{i} E_j +\nabla_{j} E_i )
+\frac{\mu^{2} }{16m^3 } (\nabla_{i} E_j +\nabla_{j} E_i )^{2}
\nonumber \\
& & +\frac{\mu^{2} }{4m^3 } (\nabla \times E)^{2}
+\left[ \frac{\mu^{2} }{8m^3 } \dot{E}^{2} + \frac{\mu^{2} }{8m^3 }
(\nabla \times B)^{2} -\frac{\mu^{2} }{4m^3 } (\nabla \times B) \cdot \dot{E}
\right] \nonumber \\
& & -\frac{\mu^{2} }{8m^3 } (\nabla \cdot E)^{2}
-\frac{\mu^{2} }{4m^3 } i\sigma \cdot (\nabla \times E) (\nabla \cdot E)
\label{hfwp0}
\end{eqnarray}
It is clear that the specification of terms to be kept given at the beginning
of section \ref{evalsec} is sufficient to guarantee that (\ref{hfwp0}) is
complete to the desired order, i.e., up to second order in the external
fields, with up to first derivatives of those fields: All terms up to
second order in the fields were kept; in the $p=0$ case, only terms
containing no $p$ operators are relevant; from among these, only terms
up to altogether fourth order in $E$, $B$, $\partial /\partial t$, $\nabla $
are needed, since each of the at most two external fields can absorb at most
one derivative.

In the form (\ref{hfwp0}), it has not yet been assumed that electric charges
or currents are absent; it includes, e.g., the $\nabla \cdot E$ structure
that embodies the influence of an external electric charge density which
enters effects such as the Darwin term. If one specializes to the vacuum,
the terms grouped in square brackets can be combined using the vacuum Maxwell
equations for the external fields,
\begin{equation}
\nabla \cdot E = 0 \ \ \ \ \ \ \ \ \ \ \nabla \cdot B = 0 \ \ \ \ \ \ \ \ \ \
\nabla \times E = -\dot{B} \ \ \ \ \ \ \ \ \ \ \nabla \times B = \dot{E}
\end{equation}
yielding
\begin{eqnarray}
W_{FW}^{p=0} &=& m -\mu \sigma \cdot B
+\frac{\mu }{2m} i\sigma \cdot \dot{B} +\frac{\mu^{2} }{2m} E^2 \nonumber \\
& & - \frac{\mu^{2} }{8m^2 } \sigma \cdot (E\times \dot{E} )
-\frac{\mu^{2} }{8m^2 } \sigma_{i} E_j (\nabla_{i} B_j +\nabla_{j} B_i )
-\frac{\mu^{2} }{4m^2 } \sigma_{i} B_j (\nabla_{i} E_j +\nabla_{j} E_i )
\label{wp0} \\
& & +\frac{\mu^{2} }{16m^3 } (\nabla_{i} E_j +\nabla_{j} E_i )^{2}
+\frac{\mu^{2} }{4m^3 } \dot{B}^{2}
\nonumber
\end{eqnarray}
The terms quadratic in the external fields can now be directly compared
to the form of the effective Hamiltonian defining the corresponding
polarizabilities. They constitute the Foldy contributions that have to
be separated out in the effective Hamiltonian to isolate the proper
polarizabilities that arise as a consequence of the extended, composite
character of the neutron. These results generalize the well-known Foldy
term that is associated specifically with the $E^2 $ structure.

\subsection{Perturbative corrections to the local limit}
\label{perturbsec}
It remains to evaluate the perturbative corrections
to (\ref{hfwp0}) or (\ref{wp0}) due to the spatially dependent terms on
the right-hand sides of (\ref{bxtcorr}) and (\ref{extcorr}). As motivated
in the preceding section, this will be accomplished in a model setting
where the neutron is placed in a box $[-L/2,L/2]^{3} $ with periodic
boundary conditions. The unperturbed Hamiltonian is taken to be the term
$H^{0} =p^2 /(2m)$ from (\ref{hfwall}). The unperturbed spatial wave
functions are
\begin{equation}
\psi_{\vec{n} }^{0} (\vec{x} )
= \frac{1}{L^{3/2} } e^{i 2\pi \vec{n} \cdot \vec{x} /L}
\end{equation}
with integer $n_i $ and energies
$W_{\vec{n} }^{0} =2\pi^{2} (\vec{n} )^2 /(mL^2 )$. For the following, it
is not necessary to specify a spin quantization axis, since the perturbations
are proportional to the unit matrix in the $p=0$ subspace; one could, e.g.,
choose to diagonalize (\ref{hfwp0}) to completely decouple the entire
Hamiltonian in that subspace. Starting with the electric field term
from (\ref{extcorr}),
\begin{equation}
H^E = \frac{\mu^{2} }{2m} \left( \nabla_{j} E_i \nabla_{k} E_i \right)
x_j x_k \ ,
\end{equation}
first-order perturbation theory is sufficient, yielding
\begin{equation}
W^{E,1} = \langle p=0 | H^E | p=0 \rangle
=(mL)^2 \frac{\mu^{2} }{24m^3 } \nabla_{j} E_i \nabla_{j} E_i \ .
\label{correctE}
\end{equation}
Note that the strength of this contribution relative to analogous terms
in (\ref{hfwp0}) or (\ref{wp0}) is controlled by the parameter $mL$,
making explicit the dependence of these perturbative corrections on
model assumptions about the environment. Additional remarks on this point
follow further below. The magnetic field term from (\ref{bxtcorr}),
\begin{equation}
H^B = -\mu \sigma_{i} \left( \nabla_{j} B_i \right) x_j
\end{equation}
on the other hand requires second-order perturbation theory (it does not
contribute at first order; indeed, $H^B |_{p=0} =0$). Summed over a complete
set of intermediate spins $S$, the moduli squared of the transition
matrix elements are
\begin{equation}
\sum_{S} \left| \langle \vec{n} ,S | H^B | p=0 \rangle \right|^{2} =
\sum_{j} \mu^{2} \left( \nabla_{j} B_i \nabla_{j} B_i \right)
\left( \frac{L}{2\pi n_j } \right)^{2} \delta_{\vec{n} , n_j \vec{e}_{j} }
\end{equation}
and hence, the resulting perturbative correction is
\begin{equation}
W^{B,2} = \sum_{\vec{n} \neq 0, S}
\frac{|\langle \vec{n} ,S | H^B | p=0 \rangle |^{2} }{W_0^0 -
W_{\vec{n} }^{0} } = -(mL)^4 \frac{\mu^{2} }{360m^3 }
\nabla_{j} B_i \nabla_{j} B_i
\label{correctB}
\end{equation}
with a strength again controlled by the parameter $mL$.

Of course, the formal limit $mL \rightarrow 0$ in which the correction
terms (\ref{correctE}) and (\ref{correctB}) vanish is not compatible
with the notion of a well-defined one-particle neutron state. A natural
lower bound is provided by equating $L$ with the Compton wavelength,
i.e, $mL=2\pi $. On the other hand, the derivative expansion employed
here is best applicable when keeping $L$ as small as possible. In typical
Lattice QCD calculations, $mL$ ranges between values of about 10-30; in
this range, the perturbative effects (\ref{correctE}) and (\ref{correctB})
are considerably stronger than the analogous terms in (\ref{hfwp0}) or
(\ref{wp0}). This also remains the case if one replaces the non-relativistic
unperturbed energies in (\ref{correctB}) with their relativistic
counterparts; at $mL=2\pi $, the result (\ref{correctB}) is then enhanced
by a factor 1.25, and the factor approaches unity with rising $mL$.
Of course, in any complete analysis of, e.g., a Lattice QCD calculation,
these perturbative corrections constitute only one among many systematic
effects that must be accounted for in order to extract physical results.

%%%%%%%%%%%%%%%%%%%%%

\section{Neutron at zero velocity}
\label{v0sec}
%%%%%%%%%%%%%%%%%%%%%
\subsection{Velocity operator}
\label{v0seca}
Since, in a gauge theory, zero momentum is not synonymous with zero
velocity, it is interesting to consider separately the energy of a neutron
at rest in the sense of being in a zero-velocity state \cite{coon1,coon2}.
The velocity operator is defined as the total time derivative of the
position operator, which can be obtained from the commutator of the
position with the Hamiltonian,
\begin{equation}
v_n = -i [x_n , H_{FW} ] = \frac{\partial H_{FW} }{\partial p_n }
\label{xdot}
\end{equation}
Note that the position operator commutes with the external fields, which
are treated as given functions of position and time.

The concept of a zero-velocity state is not unproblematic. The starting
point of the following discussion is the assumption that there is a state
$|v=0\rangle $ that is annihilated by all components of the velocity
operator, $v_n |v=0\rangle =0$, and explore how far this assumption
carries. As one might already suspect from (\ref{xdot}), and as will
become more clear from the explicit expressions below, the individual
components $v_n $ do not in general commute, and therefore one cannot in
general expect to construct simultaneous eigenstates of these operators.
However, this by itself does not exclude the possibility of a state with,
specifically, $v_n |v=0\rangle =0$, similar to the case of the standard
angular momentum algebra, for which one does have $J_n |j=0\rangle =0$.
No attempt will be made to construct a state $|v=0\rangle $ explicitly.
Rather, to the extent that such a state exists, the results obtained
will apply to it.

Certainly, although the concept of a zero-velocity neutron is a priori
physically meaningful, such a state can in general only be of a transitory
nature in the presence of external electromagnetic fields.
Already at the classical level, a magnetic moment is accelerated
in an inhomogeneous magnetic field. In that case, one can only expect a
consistent treatment in terms of such a state up to a limited order in
a derivative expansion such as pursued in this work. To the order
considered here, however, no inconsistency related to the inhomogeneity
of the magnetic field will become apparent.

On the other hand, a zero-velocity quantum state presumably requires a
certain spatial extension, and, as already discussed in previous sections,
a derivative expansion of the external fields can only capture a limited
range of spatial behavior. This likewise may manifest itself in
inconsistencies beyond a certain order in the derivative expansion.
Such an inconsistency will in fact emerge in the treatment below,
specifically with respect to the terms in the neutron mass proportional
to $\nabla_{i} E_j \nabla_{i} E_j $ and $(\nabla \cdot E)^2 $. The
coefficients of these terms will be seen to be ambiguous, and no
completely cogent scheme has become apparent within the calculational
framework employed here that would permit a definite determination of
these terms. Presumably, this would require going beyond the derivative
expansion (\ref{heff}) from the very beginning.

An alternative definition of a zero-velocity state can be contemplated
in which one considers only the velocity squared, and one assumes only
the existence of a state $|v^2 =0\rangle $ with $v^2 |v^2 =0\rangle =0$.
This will likewise be explored further below; while it does circumvent the
issue of the individual components $v_n $ not commuting, this approach
has its own difficulties and does not finally resolve the aforementioned
ambiguity.

In \cite{coon2}, a definition of a zero-velocity state is considered
for which only the considerably weaker condition $\langle v \rangle =0$
is posited. Such a treatment requires a construction of the state, which
is beyond the scope of the present work. It should be noted, however,
that also this definition does not eliminate the transitory nature of
a zero-velocity state. Taken by by itself, this condition would encompass
generic bound states, presumably including highly excited ones that may be
interpreted in terms of classical orbits. In that limit, it becomes
questionable whether the neutron can still be viewed as being truly at
rest, and it may be necessary to supply a supplementary characterization
of the states to be considered.

For $H_{FW} $ given by (\ref{hfwall}), the velocity operator takes the
form
\begin{equation}
v_n = \frac{1}{m} p_n - \frac{1}{2m^3 } p^2 p_n + G_n
\label{vdef}
\end{equation}
where $G_n $ summarizes all the terms that are at least linear in the
external fields,
\begin{eqnarray}
G_n &=& -\frac{\mu }{m} (\sigma \times E)_{n}
+\frac{\mu }{4m^2 } \left( -\dot{E}_{n} + (\nabla \times B)_{n}
-i\sigma_{i} \nabla_{i} B_n \right)
+\frac{\mu^{2} }{2m^3 } \left(
-\sigma_{j} E_i \nabla_{j} E_k \epsilon_{ikn}
+(\sigma \cdot E) (\nabla \times E)_{n} \right. \nonumber \\
& & \left. +3i E_i \nabla_{n} E_i -i E_i \nabla_{i} E_n
-(\nabla \cdot E) (\sigma \times E)_{n} -i (\nabla \cdot E) E_n
-iB_i \nabla_{i} B_n +\sigma \cdot (\nabla \times B) B_n
-(\sigma \cdot \dot{E}) B_n \right) \nonumber \\
& & +\frac{\mu }{2m^3 } \left( -i \sigma_{j} \epsilon_{jin} \nabla_{k} E_i p_k
-i \sigma_{k} \epsilon_{kij} \nabla_{n} E_i p_j
+i \sigma \cdot (\nabla \times E) p_n
+(\nabla \cdot E) p_n \right) \nonumber \\
& & +\frac{\mu }{2m^2 } \left( \sigma_{i} B_n p_i +\sigma_{n} B\cdot p
\right) + \frac{\mu }{2m^3 } \left(
2\sigma \cdot (E \times p) p_n +(\sigma \times E)_{n} p^2 \right)
\label{collectgn}
\end{eqnarray}

\subsection{Action of Foldy-Wouthuysen Hamiltonian on $|v=0\rangle $
state}
To derive the action of $H_{FW} $ on a zero-velocity state $|v=0\rangle $,
the momentum operator can be eliminated in favor of the velocity operator
using the following iterative scheme. Rearranging (\ref{vdef}) into
\begin{equation}
p_n = \left( 1-\frac{p^2 }{2m^2 } \right)^{-1} m (v_n -G_n )
= \left( 1+\frac{p^2 }{2m^2 } + \ldots \right) m (v_n -G_n )
\end{equation}
where higher orders in $p$ can be neglected in the right-hand expression,
it follows that the action of $p_n $ on a zero-velocity state $|v=0\rangle $
yields
\begin{equation}
p_n = -m\left( 1+\frac{p^2 }{2m^2 } \right) G_n
\label{presolv}
\end{equation}
thus generating at least one power of an external field. With $H_{FW} $
given by (\ref{hfwall}),
i.e., with all momentum operators commuted through to the right, one can
let the right-most momentum operator act on the zero-velocity state and
therefore substitute it with (\ref{presolv}). This yields an expression
in which the momentum operators are not all ordered to the right; one
then has to commute all momentum operators to the right again. At that
point, one can iterate the procedure, i.e., again let the right-most
momentum operator act on the zero-velocity state, etc. This procedure
has to be performed at most twice, since every iteration generates an
additional power of an external field.

Before proceeding, it is convenient at this point to revisit the
specification given at the beginning of section \ref{evalsec} of the
terms that were to be kept in deriving the Foldy-Wouthuysen Hamiltonian
(\ref{hfwall}). Consider first terms quadratic in the external fields.
To begin with, as far as $H_{FW} $ itself is concerned, no additional
factor of $p$ (which, as always, would be ordered to the right) needs
to be taken into account, since such a factor of $p$, applied to the
zero-velocity state, would generate an additional external field.
However, one must also ensure that terms in $G_n $ are kept to
sufficiently high order; deriving $G_n $ from $H_{FW} $ removes one
factor of $p$. As a result, also terms in $H_{FW} $ quadratic in the
external fields and linear in $p$ must be kept, since the corresponding
term in $G_n $ is then simply quadratic in the external fields with no
further factors of $p$. However, if there were yet one more factor
of $p$, then $G_n $ would be quadratic in the external fields with an
additional factor $p$ which, acting on the zero-velocity state, would
generate an additional external field. Such terms can therefore be
discarded.

Similarly, consider terms linear in the external fields. As far as
$H_{FW} $ itself is concerned, only two factors of $p$ need to be
taken into account; the rightmost of them generates a second external
field when applied to the zero-velocity state, and this external field
can absorb the other factor of $p$ as a derivative. However, any
additional factor $p$ would commute through to the right and generate
a third external field when applied to the zero-velocity state. Again,
however, one has to also ensure that sufficiently many terms in $G_n $
are kept. Keeping three powers of $p$ in $H_{FW} $ corresponds to
having two powers of $p$ in $G_n $; the rightmost one generates a
second external field, which can absorb the other factor of $p$ as a
derivative, rendering $G_n $ simply quadratic in the external fields
with no further factors of $p$. However, if there were yet one more factor
of $p$, then it would commute through to the right and generate a third
external field.

Finally, consider the overall order in
$E$, $B$, $\partial /\partial t$, $\nabla $, $p$. Certainly, as far as
$H_{FW} $ itself is concerned, altogether fourth order is sufficient:
Every factor $p$ generates either an external field or a derivative;
the desired limitation to at most two external fields, which each can absorb
at most one derivative, implies that only terms up to fourth order in these
objects collectively are relevant. Now, keeping terms up to fourth
order in $H_{FW} $ corresponds to consistently keeping only terms up to
third order in $G_n $. This is nevertheless sufficient, because in $H_{FW} $,
any rightmost factor $p$ always comes multiplied with external fields or
further factors of $p$. Thus, after applying said rightmost factor $p$ to
the zero-velocity state, generating a factor $G_n $, that factor $G_n $
always comes multiplied with another factor of $p$ or an external field.
Therefore, the fourth order terms in $G_n $ are in fact irrelevant and
it is indeed sufficient to construct $G_n $ consistently only to third
order. Thus, in $H_{FW} $ only altogether fourth order in
$E$, $B$, $\partial /\partial t$, $\nabla $, $p$ is required.

Continuing with the above scheme of eliminating the momentum operator
to derive the action of $H_{FW} $ on a zero-velocity state, note that
the identification (\ref{presolv}) in general introduces an
ambiguity: Since the operators $p_i $ and $G_j $ don't commute, the order
in which one applies momentum operators can make a difference, i.e., in
general it can happen that, effectively, $[p_i ,p_j ]\neq 0$. However,
zero-velocity states on which this occurs presumably should not be
viewed as acceptable solutions of the present small-field perturbative
expansion. If $[p_i ,p_j ]\neq 0$ when applied to a wave function, this
implies that the wave function contains singular vorticity. It thus differs
strongly from zero-momentum states, which are spatially constant. However,
the present small-field expansion presupposes that zero-velocity states
are perturbations of zero-momentum states, with corrections suppressed by
powers of the small external fields. For zero external field, zero momentum
and zero velocity coincide. Thus, states with singular vorticity that yield
$[p_i ,p_j ]\neq 0$ ought to be excluded from consideration a priori in
order to preserve a consistent perturbative small-field expansion
scheme. Thus, the requirements for a consistent treatment would appear
to include treating products of momentum operators as
\begin{equation}
p_i p_j = \frac{1}{2} \{ p_i , p_j \} + \frac{1}{2} [ p_i , p_j ]
\label{symme}
\end{equation}
and setting $[p_i ,p_j ]=0$. This working assumption will be explored
further below and a concluding critique will be given in section \ref{incons}.

To exhibit clearly the emergence of the ambiguities discussed above,
it is useful to proceed as follows. Consider, to begin with, the
operator $p^2 = p_n p_n $ appearing in $H_{FW} $. In this operator, no
ordering ambiguity arises regarding an initial application of the
identification (\ref{presolv}). Multiplying (\ref{presolv}) from the
left with $p_n $ and commuting all momentum operators through to the
right yields (as above, in the following, $\nabla \cdot B =0$ will be
dropped by virtue of Maxwell's equations for the external fields)
\begin{eqnarray}
p^2 &=& i\mu \sigma \cdot (\nabla \times E) +\mu \sigma \cdot (E \times p)
\nonumber \\
& & +\frac{\mu^{2} }{2m^2 } \left( -3 \nabla_{i} E_j \nabla_{i} E_j
+\nabla_{i} E_j \nabla_{j} E_i
+i\sigma \cdot (\nabla \times E) \nabla \cdot E + (\nabla \cdot E)^2
+\nabla_{i} B_j \nabla_{j} B_i \right) \nonumber \\
& & +\frac{\mu }{2m^2 } \left(
2i\epsilon_{ijk} \sigma_{k} \nabla_{l} E_i p_l p_j
-i\sigma \cdot (\nabla \times E) p^2 -\nabla \cdot E p^2 \right) \nonumber \\
& & +\frac{\mu }{4m} \left( \dot{E} \cdot p -(\nabla \times B)\cdot p
+3i\sigma_{i} \nabla_{i} B_j p_j -4\sigma_{j} B_i p_j p_i \right)
\label{p2}
\end{eqnarray}
In this form, one now observes potential ordering ambiguities in the
third and fourth lines. However, instead of contemplating further
manipulations of this form on its own, one can proceed by inserting it
into the full Hamiltonian $H_{FW} $, upon which one observes several
cancellations of these potential ambiguities,
\begin{eqnarray}
H_{FW} &=& m -\mu \sigma \cdot B -\frac{\mu }{2m} \nabla \cdot E
+\frac{\mu^{2} }{2m} E^2 \nonumber \\
& & + \frac{\mu^{2} }{8m^2 } \left(
2\sigma \cdot (E\times \dot{E} ) -3 \sigma \cdot (E\times (\nabla \times B))
-\sigma_{i} E_j (\nabla_{i} B_j +\nabla_{j} B_i )
-2\sigma_{i} B_j (\nabla_{i} E_j +\nabla_{j} E_i ) \right) \nonumber \\
& & + \frac{\mu^{2} }{32m^3 } \left(
4\dot{E}^{2} -8(\nabla \times B) \cdot \dot{E} +2(\nabla \times B)^2
+ (\nabla_{i} B_j +\nabla_{j} B_i )^2 \right. \nonumber \\
& & \left. \hspace{3cm}
-12 \nabla_{i} E_j \nabla_{i} E_j +4\nabla_{i} E_j \nabla_{j} E_i
+4(\nabla \cdot E)^2 \right) \nonumber \\
& & -\frac{\mu }{2m} \sigma \cdot (E\times p)
+\frac{\mu }{8m^2 } \left( (\nabla \times B)\cdot p -\dot{E} \cdot p
+i\sigma_{i} \nabla_{i} B_j p_j \right) \nonumber \\
& & -\frac{1}{8m^3 } p^4
\label{hv0}
\end{eqnarray}
Fortuitously, only one ambiguous term remains, namely the term proportional
to $p^4 $, which will be considered separately below. Aside from that term,
there are only ones containing at most one power of momentum, and therefore
no ordering ambiguities. Since the terms containing one power of momentum
already exhibit one power of the external fields, only a few of the terms
appearing in (\ref{presolv}) in conjunction with (\ref{collectgn}) remain
relevant when eliminating that momentum operator; the other terms only
contribute at the third order in the external fields or higher. Namely,
only the first four terms in (\ref{collectgn}) must be retained, and also
the $p^2 $ operator in the parenthesis in (\ref{presolv}) can be dropped.
Carrying out this elimination yields
\begin{eqnarray}
H_{FW} &=& m -\mu \sigma \cdot B -\frac{\mu }{2m} \nabla \cdot E
-\frac{\mu^{2} }{2m} E^2
+\frac{\mu^{2} }{64m^3 } \left( 6\dot{E}^{2} -3(\nabla \times B)^2
-12(\nabla \times B) \cdot \dot{E} \right) \nonumber \\
& & +\frac{7\mu^{2} }{128m^3 } (\nabla_{i} B_j +\nabla_{j} B_i )^2
+\frac{\mu^{2} }{8m^3 } \left( (\nabla \cdot E)^2
-3 \nabla_{i} E_j \nabla_{i} E_j +\nabla_{i} E_j \nabla_{j} E_i \right)
\nonumber \\
& & -\frac{\mu^{2} }{4m^2 } \left(
\sigma \cdot (E\times (\nabla \times B))
+\sigma_{i} B_j (\nabla_{i} E_j +\nabla_{j} E_i ) \right) \nonumber \\
& & -\frac{1}{8m^3 } p^4
\end{eqnarray}
It remains to treat the operator $p^4 $. Multiplying the form (\ref{p2})
for the operator $p^2 $ by another factor $p^2 $ from the left, one may
discard the majority of the terms because they only contribute at orders
that are dropped in the expansion pursued here. The remaining relevant
terms are
\begin{equation}
p^4 = p^2 \left[
i\mu \sigma \cdot (\nabla \times E) +\mu \sigma \cdot (E \times p) \right]
\label{p4init}
\end{equation}
Two ways of proceeding suggest themselves: On the one hand, one may
view $p^4 $ as the successive application of two $p^2 $ operators,
i.e., first fully resolve the term in the square brackets by applying
the remaining momentum operator to the zero-velocity state; on the
other hand, one may follow the general scheme laid out above, commuting
all momentum operators to the right in (\ref{p4init}) immediately, and
then treating them in a symmetrized fashion as suggested by (\ref{symme}).
Note that the fully symmetrized form
$p^4 = (1/3) (p_i p_i p_j p_j +p_i p_j p_i p_j + p_i p_j p_j p_i )$
corresponds to a weighted average of these two alternatives.

Starting with the former procedure, by inserting (\ref{presolv}) and only
keeping relevant terms, one obtains
\begin{eqnarray}
p^4 &=& p^2 \left[ i\mu \sigma \cdot (\nabla \times E) +
\mu^{2} \sigma \cdot (E \times (\sigma \times E)) \right] \\
&=& p^2 \left[ i\mu \sigma \cdot (\nabla \times E) + 2 \mu^{2} E^2 \right] \\
&=& i\mu \sigma \cdot (\nabla \times E) p^2
-4\mu^{2} \nabla_{i} E_j \nabla_{i} E_j \\
&=& \left[ i\mu \sigma \cdot (\nabla \times E) \right]^2
-4\mu^{2} \nabla_{i} E_j \nabla_{i} E_j \\
&=& -5 \mu^{2} \nabla_{i} E_j \nabla_{i} E_j
+\mu^{2} \nabla_{i} E_j \nabla_{j} E_i
\label{p4p2p2}
\end{eqnarray}
On the other hand, immediately commuting momentum operators to the
right in (\ref{p4init}), one has
\begin{equation}
p^4 = i\mu \sigma \cdot (\nabla \times E) p^2
-i\mu \epsilon_{ijk} \sigma_{i} \nabla_{l} E_j \{ p_l , p_k \}
\end{equation}
exhibiting the ambiguity in ordering momentum operators in the last
term, treated as suggested by (\ref{symme}). Again inserting
(\ref{presolv}) and only keeping relevant terms, one has
\begin{equation}
p^4 = \mu^{2} (\nabla \cdot E)^2
-4 \mu^{2} \nabla_{i} E_j \nabla_{i} E_j
+\mu^{2} \nabla_{i} E_j \nabla_{j} E_i
\label{p4commut}
\end{equation}
having used $(\nabla \times E)_{j} (\nabla_{i} E_j -\nabla_{j} E_i )=0$.
The expressions (\ref{p4p2p2}) and (\ref{p4commut}) disagree; evidently,
varying the ordering of momentum operators corresponds to trading off
terms $(\nabla \cdot E)^2 $ and $-\nabla_{i} E_j \nabla_{i} E_j $ in the
Foldy-Wouthuysen Hamiltonian. The coefficients of these terms in $H_{FW} $
are ambiguous unless a cogent rationale for choosing a particular
ordering can be constructed. This will be revisited in section \ref{incons}.
The term $\nabla_{i} E_j \nabla_{j} E_i $ does appear to be determined
with a unique coefficient, such that the $\nabla_{i} E_j \nabla_{j} E_i $
contribution completely cancels in $H_{FW} $. With the ambiguous terms
unresolved, the energy of the neutron in a zero-velocity state
$|v=0\rangle $ takes the form
\begin{eqnarray}
W_{FW}^{v=0} &=&
m -\mu \sigma \cdot B -\frac{\mu }{2m} \nabla \cdot E
-\frac{\mu^{2} }{2m} E^2
+\frac{\mu^{2} }{64m^3 } \left[ 6\dot{E}^{2} -3(\nabla \times B)^2
-12(\nabla \times B) \cdot \dot{E} \right] \nonumber \\
& & +\frac{7\mu^{2} }{128m^3 } (\nabla_{i} B_j +\nabla_{j} B_i )^2
-\frac{\mu^{2} }{4m^2 } \left(
\sigma \cdot (E\times (\nabla \times B))
+\sigma_{i} B_j (\nabla_{i} E_j +\nabla_{j} E_i ) \right) \nonumber \\
& & +O\left( (\nabla \cdot E)^2 \right)
+O\left( \nabla_{i} E_j \nabla_{i} E_j \right)
\end{eqnarray}
Specializing to the vacuum and using the vacuum Maxwell equations for the
external fields, which allows one to combine the terms grouped in the square
brackets, one arrives at
\begin{eqnarray}
W_{FW}^{v=0} &=&
m -\mu \sigma \cdot B
-\frac{\mu^{2} }{2m} E^2 -\frac{9\mu^{2} }{64m^3 } \dot{E}^2
+\frac{7\mu^{2} }{128m^3 } (\nabla_{i} B_j +\nabla_{j} B_i )^2 \nonumber \\
& & -\frac{\mu^{2} }{4m^2 } \sigma \cdot (E\times \dot{E})
-\frac{\mu^{2} }{4m^2 } \sigma_{i} B_j (\nabla_{i} E_j +\nabla_{j} E_i )
\label{wv0} \\
& & +O\left( \left( \nabla_{i} E_j + \nabla_{j} E_i \right)^{2} \right)
+O\left( \dot{B}^{2} \right)
\nonumber
\end{eqnarray}
taking into account the decomposition $\nabla_{i} E_j \nabla_{i} E_j =
(1/4) \left[ ( \nabla_{i} E_j + \nabla_{j} E_i )^2 +
( \nabla_{i} E_j - \nabla_{j} E_i )^2 \right] $ as well as
$( \nabla_{i} E_j - \nabla_{j} E_i )^2 = 2(\nabla \times E)^2
=2\dot{B}^{2} $.
Compared to the $p=0$ case, the $E^2 $ term changes sign, as has been
previously observed in \cite{coon1}. Also the higher order terms are
modified substantially, some disappearing entirely and new ones
appearing. Two terms remain undetermined in the $v=0$ case.

\subsection{Action of Foldy-Wouthuysen Hamiltonian on $|v^2 =0\rangle $
state}
Before considering the relative merits of the different schemes of treating
the ordering of momentum operators exhibited in the previous section, it
is useful to also have at hand the expressions resulting when acting on a
$|v^2 =0\rangle $ state, for which only the property $v^2 |v^2 =0\rangle =0$
is assumed. Proceeding in analogy to the argument leading to
eq.~(\ref{presolv}), taking the square of eq.~(\ref{vdef}) yields
\begin{equation}
v^2 = \frac{p^2 }{m^2 } \left( 1- \frac{p^2 }{m^2 } \right) + F
\end{equation}
with $F$ summarizing all the terms that are at least linear in the
external fields,
\begin{equation}
F = \frac{p_n }{m} \left( 1- \frac{p^2 }{2m^2 } \right) G_n
+ G_n \frac{p_n }{m} \left( 1- \frac{p^2 }{2m^2 } \right) +G_n G_n
\end{equation}
This can be rearranged to construct an iterative scheme for eliminating
$p^2 $ in favor of $v^2 $,
\begin{equation}
p^2 = m^2 \left( 1- \frac{p^2 }{m^2 } \right)^{-1} \left( v^2 -F \right)
= m^2 \left( 1+ \frac{p^2 }{m^2 } + \ldots \right) \left( v^2 -F \right)
\end{equation}
and thus, applied to a $|v^2 =0\rangle $ state,
\begin{eqnarray}
p^2 &=& -(m^2 +p^2 ) F \\
&=& -mp_n \left( 1+ \frac{p^2 }{2m^2 } \right) G_n
-m \left( 1+ \frac{p^2 }{m^2 } \right) G_n p_n - (m^2 +p^2 ) G_n G_n
\label{p2v2}
\end{eqnarray}
Note that the first term in (\ref{p2v2})  corresponds to
eq.~(\ref{presolv}) with an extra $p_n $ applied from the left; i.e.,
the expression for $p^2 $ obtained here, acting on a $|v^2 =0\rangle $
state, differs from the one obtained when acting on a $|v=0\rangle $ state,
cf.~(\ref{p2}), by the two additional terms in (\ref{p2v2}). For the
discussion in the next section, it is useful to observe that these
additional terms can also be cast as follows,
\begin{eqnarray}
p^2 &=& -mp_n \left( 1+ \frac{p^2 }{2m^2 } \right) G_n
-(m^2 +p^2 ) G_n v_n
\label{p2v0}
\end{eqnarray}
as can be verified by inserting (\ref{vdef}), again discarding terms that
are of too high order. Note that any $|v=0\rangle $ state is also a
$|v^2 =0\rangle $ state, and therefore the expressions for $p^2 $ derived
in the two cases must be consistent with one another when applied to
$|v=0\rangle $ states. Indeed, the additional terms vanish when one
uses $v_n |v=0\rangle =0$. On the other hand, for $|v^2 =0\rangle $ states
that are not also known to be $|v=0\rangle $ states, there is no a priori
guarantee that the additional terms vanish.

To assemble the Foldy-Wouthuysen Hamiltonian acting on $|v^2 =0\rangle $
states, one can reuse the result (\ref{hv0}), merely supplementing it
with the additional terms entering $p^2 $ (multiplied by the appropriate
prefactor, $1/(2m)$). Using (\ref{collectgn}), commuting momentum operators
to the right, and dropping terms that are of too high order, these terms read
\begin{eqnarray}
-mG_n p_n &=& \mu (\sigma \times E) \cdot p
-\frac{\mu }{4m} \left( -\dot{E} \cdot p + (\nabla \times B) \cdot p
-i\sigma_{i} \nabla_{i} B_n p_n \right) \nonumber \\
& & -\frac{\mu }{2m^2 } \left( -2i \sigma_{j} \epsilon_{jin}
\nabla_{k} E_i p_k p_n
+i \sigma \cdot (\nabla \times E) p^2 +(\nabla \cdot E) p^2 \right)
-\frac{\mu }{m} \sigma_{i} B_n p_i p_n \\
-\frac{p^2 }{m} G_n p_n &=& -2i \frac{\mu }{m^2 } \epsilon_{ijk}
\sigma_{j} \nabla_{l} E_k p_l p_i \\
-m^2 G_n G_n &=& -m^2 \left(
-\frac{\mu }{m} \sigma \times E
+\frac{\mu }{4m^2 } \left( -\dot{E} + \nabla \times B
-i\sigma_{i} \nabla_{i} B \right) \right)^{2} \nonumber \\
& & +\frac{\mu^{2} }{2m^2 }
\left( -4 \nabla_{i} E_j \nabla_{i} E_j + \nabla_{i} E_j \nabla_{j} E_i
+ (\nabla \cdot E)^2 + (\nabla \cdot E) i\sigma \cdot (\nabla \times E)
\right) \nonumber \\
& & +\frac{\mu^{2} }{2m}
\left( 3\sigma_{j} B_i \nabla_{i} E_j - (\nabla \cdot E) (\sigma \cdot B)
-iB\cdot (\nabla \times E) \right) \\
-p^2 G_n G_n &=& 4\frac{\mu^{2} }{m^2 } \nabla_{i} E_j \nabla_{i} E_j
\end{eqnarray}
In addition, in the operator $p^4 = p^2 p^2 $, one can likewise substitute
the right-hand $p^2 $ with (\ref{p2v2}); after commuting the other $p^2 $
operator to the right and discarding terms of too high order, the remaining
terms are
\begin{equation}
p^4 = i\mu \sigma \cdot (\nabla \times E) p^2
-4i\mu \epsilon_{ijk} \sigma_{i} \nabla_{l} E_j p_k p_l
+4\mu^{2} \nabla_{i} E_j \nabla_{i} E_j
\end{equation}
Consider, to begin with, the terms now appearing in $H_{FW} $ that are
linear in $B$ and contain no $E$. These read
\begin{equation}
\left. H_{FW} \right|_{E=0,\ \mbox{\scriptsize linear in } B } \ =
-\mu \sigma \cdot B + \frac{\mu }{4m^2 } \left(
i\sigma_{i} \nabla_{i} B_j p_j -2 \sigma_{i} B_j p_i p_j \right)
\end{equation}
This form implies that one reaches an impasse in the treatment of
external magnetic fields; the present scheme of acting on a
$|v^2 =0\rangle $ state only allows one to eliminate the operator
$p^2 $, but not individual components $p_n $. The above terms thus
cannot be evaluated any further. The $|v^2 =0\rangle $ state scheme
therefore has the significant drawback of not permitting a well-defined
treatment of external magnetic fields. In the following, only the case
$B=0$ will therefore be considered.

Fortuitously, when one assembles the terms containing electric fields,
all problematic terms cancel and one is left at most with additional
factors $p^2 $ to resolve. Note that, when already multiplied by one power
of an external field from the left, the only term in $p^2 $ that remains
relevant to the desired order is $i\mu \sigma \cdot (\nabla \times E)$.
Carrying out the remaining evaluation, one arrives at the Foldy-Wouthuysen
Hamiltonian
\begin{equation}
\left. H_{FW} \right|_{B=0} = m -\frac{\mu }{2m} \nabla \cdot E
-\frac{\mu^{2} }{2m} E^2 +\frac{3\mu^{2} }{32m^3 } \dot{E}^{2}
+\frac{3\mu^{2} }{8m^3 } (\nabla \cdot E)^2
+\frac{\mu^{2} }{2m^3 } \nabla_{i} E_j \nabla_{i} E_j
\end{equation}
This reproduces the form obtained for $|v=0\rangle $ states, except for
the ambiguous $(\nabla \cdot E)^2 $ and $\nabla_{i} E_j \nabla_{i} E_j $
terms, which appear here with yet different coefficients than in either of
the alternative schemes discussed in the previous section.

\subsection{Inconsistencies in defining a $|v=0\rangle $ state at
order $O(\nabla E \nabla E) $}
\label{incons}
As has been argued already further above, a consistent zero-velocity state
$|v=0\rangle $ ought to satisfy \linebreak $[p_i ,p_j ] |v=0\rangle =0$
in order to represent a bona fide small-field perturbation of a
zero-momentum state. This condition was used to resolve the ordering
ambiguity in products of momentum operators, cf.~(\ref{symme});
whenever a product of momentum operators acts on a zero-velocity
state, the product is to be symmetrized.

However, this prescription cannot be consistently maintained under
all circumstances. Recalling the discussion of eqs.~(\ref{p2v2}) and
(\ref{p2v0}), since any $|v=0\rangle $ state is also a $|v^2 =0\rangle $
state, the expression (\ref{p2v2}) for the operator $p^2 $ acting on a
$|v^2 =0\rangle $ state must equal the expression obtained when acting
on a $|v=0\rangle $ state, which only includes the first term in (\ref{p2v2}).
Indeed, as exhibited in eq.~(\ref{p2v0}), the additional terms vanish
on $|v=0\rangle $ states, since they can be written in terms of an
operator that includes a factor $v_n $ on the right. This, however,
supposes a definite ordering of operators that is inconsistent with
a symmetrization prescription: In the product
\begin{equation}
G_n v_n = G_n \left( \frac{p_n }{m} + G_n \right)
\end{equation}
the operator $p_n $ must be kept to the right of any momentum operators
appearing in the $G_n $ outside of the parentheses. If one instead
were to symmetrize the product of $p_n $ with momentum operators occurring
to its left, one would alter the product $G_n v_n $, which vanishes when
acting on $|v=0\rangle $ states, into a different operator $\Gamma $ that
is not anymore a product of two factors $G_n $ and $v_n $, but represents
a new composite operator that does not vanish when acting on $|v=0\rangle $
states. One could, in effect, come to the absurd conclusion that
\begin{equation}
0 = G_n v_n |v=0\rangle = G_n \left( \frac{p_n }{m} + G_n \right) |v=0\rangle
=: \Gamma |v=0\rangle \neq 0
\end{equation}
where the non-vanishing terms again involve $(\nabla \cdot E)^2 $ and
$\nabla_{i} E_j \nabla_{i} E_j $. It appears, therefore, that, once one
attempts to determine its energy to an accuracy including the order
$O(\nabla E \nabla E) $, there exists no $|v=0\rangle $ state consistently
defined to that accuracy. One may speculate that this is a signature of a
conflict between the necessarily extended nature of a zero-velocity
state in space and the limited spatial range of a description in terms of
a derivative expansion. Ultimately, it does not appear feasible within
the present framework to determine unambiguously contributions to the
energy of a neutron at rest that are proportional to
$(\nabla \cdot E)^2 $ and $\nabla_{i} E_j \nabla_{i} E_j $. It
remains unclear whether retreating to a calculational scheme based
on $|v^2 =0\rangle $ states can provide a resolution of the ambiguities
associated with these terms, or whether it merely hides them through its
relative inflexibility in exploring different operator orderings. In any
case, the $|v^2 =0\rangle $ scheme does not allow one to treat magnetic
background fields, as seen in the previous section.

%%%%%%%%%%%%%%%%%%%%%

\section{Conclusions}
%%%%%%%%%%%%%%%%%%%%%
By constructing an appropriate Foldy-Wouthuysen transformation, the
energy of a point-like neutron in an external electromagnetic field
was determined in a combined expansion in powers of the external
field and derivatives thereof. Both the case of a zero-momentum state
as well as the case of a zero-velocity state were considered, leading
to the results (\ref{wp0}) and (\ref{wv0}), respectively, in the absence
of external charges or currents. The obtained terms mirror the ones
appearing in the effective Hamiltonian (\ref{heff}), mimicking the
effects of the polarizabilities defined there. This generalizes the
long-known result of Foldy \cite{foldy} pertaining specifically to the
dipole electric polarizability $\alpha_{E} $. In order to separate the
dipole electric polarizability proper, i.e., the effect of an actual
structural deformation of an extended neutron, from the energy shift
already experienced by a pointlike neutron, the contribution
$\alpha_{E}^{Foldy} = -\mu^{2}/m $ (in a zero-momentum state) must be
subtracted from the coefficient $\alpha_{E} $ found in the effective
Hamiltonian (\ref{heff}). Comparing the zero-momentum result (\ref{wp0})
with (\ref{heff}), the ten Foldy contributions for a zero-momentum neutron
are
\begin{equation}
\begin{array}{ccccc}
\alpha_{E}^{Foldy,p=0}=-\frac{\mu^2}{m}, &
\beta_{M}^{Foldy,p=0}=0, &
\alpha_{E\nu}^{Foldy,p=0}=0, &
\beta_{M\nu}^{Foldy,p=0}=-\frac{\mu^2}{2m^3}, &
\alpha_{E2}^{Foldy,p=0}=-\frac{3\mu^2}{m^3},\\
 & & & & \\
\beta_{M2}^{Foldy,p=0}=0, &
\gamma_{E1}^{Foldy,p=0}=\frac{\mu^2}{4m^2}, &
\gamma_{M1}^{Foldy,p=0}=0, &
\gamma_{E2}^{Foldy,p=0}=-\frac{\mu^2}{2m^2}, &
\gamma_{M2}^{Foldy,p=0}=\frac{\mu^2}{4m^2} \ .
\end{array}
\label{p0result}
\end{equation}
On the other hand, eliminating the momentum operator in favor of the
velocity operator, one can construct analogous contributions in the
zero-velocity case. In the course of this construction, operator-ordering
ambiguities were encountered that appear to signal an inconsistency in
defining a zero-velocity state to order $O(\nabla E \nabla E)$. This is
presumably due to a conflict between the spatially extended nature of
such a state and the limitation of a derivative expansion in capturing
the associated spatial behavior. The ambiguities specifically arise in
the treatment of the $p^4 $ term representing a relativistic correction
to the kinetic energy. They are, therefore, a relativistic effect, but
understanding these relativistic effects is evidently necessary for a
proper treatment of $O(\nabla E \nabla E)$ terms. These ambiguities
precluded a determination of the Foldy contributions for $\alpha_{E2} $ and
$\beta_{M\nu} $. It should be emphasized that, in view of the exhaustive
treatment presented in section \ref{v0sec} of the consequences of assuming
the existence of a zero-velocity state, the ambiguities appear to be a
consequence already of the initial assumption of the neutron energy being
represented by an expansion of the form (\ref{heff}). To remove them
would require abandoning the form (\ref{heff}) rather than merely
improving upon the calculational scheme employed here. The study of forms
more general than (\ref{heff}) lies beyond the scope of this work.
Comparing the result (\ref{wv0}) with with (\ref{heff}),
the remaining eight Foldy contributions for a zero-velocity neutron are
\begin{equation}
\begin{array}{ccccc}
\alpha_{E}^{Foldy,v=0}=\frac{\mu^2}{m}, &
\beta_{M}^{Foldy,v=0}=0, &
\alpha_{E\nu}^{Foldy,v=0}=\frac{9\mu^2}{32m^3}, &
\beta_{M2}^{Foldy,v=0}=-\frac{21\mu^2}{8m^3},\\
 & & & & \\
\gamma_{E1}^{Foldy,v=0}=\frac{\mu^2}{2m^2}, &
\gamma_{M1}^{Foldy,v=0}=0, &
\gamma_{E2}^{Foldy,v=0}=-\frac{\mu^2}{2m^2}, &
\gamma_{M2}^{Foldy,v=0}=0 \ .
\end{array}
\label{v0result}
\end{equation}
As already noted in \cite{coon1}, the sign of the Foldy contribution
for $\alpha_{E} $ is inverted going from the zero-momentum to the
zero-velocity case. Also the majority of the other contributions changes.

An application of these results that suggests itself is the analysis of
Lattice QCD calculations of hadron mass shifts in the presence of external
electromagnetic fields. Which of the scenaria investigated here is
relevant for such an analysis depends sensitively on the details of the
lattice setup; there is no simple, unique prescription for disentangling
the mass shift due to the genuine polarization of the hadron's internal
structure from the Foldy-type contributions discussed here, additional
perturbative effects such as the ones exhibited in section \ref{perturbsec},
and a variety of other systematic corrections.

The calculational scheme employed in \cite{elpol,polprog,spinpol} preserves
spatial translational invariance and thereby is able to place the neutron
in a zero-momentum state. To correct for Foldy-type effects, the $p=0$
results (\ref{p0result}) are therefore relevant; in addition, the
perturbative corrections discussed in section \ref{perturbsec} apply.
It should be noted that this relative simplicity of the spatial setup
comes at the price of using explicitly time-dependent external
electromagnetic gauge fields that induce a complicated non-stationary
time evolution; the extraction of the neutron mass shift requires a
delicate analysis of the resulting transitory neutron state.

In order to avoid such complications, the scheme employed in
\cite{alee,leea,gwgamma,lujan1,freeman,lujan2} instead uses Dirichlet
boundary conditions for the valence quarks in some space-time directions,
which permits one to introduce arbitrarily weak, temporally constant
external gauge fields. However, one consequence of using Dirichlet
boundary conditions is that the hadron cannot be in a zero-momentum
state, but instead acquires a nontrivial spatial wave function with
typical momenta of the order of $\pi /L$ (where $L$ denotes the spatial
extent of the lattice), including also distortions of the hadron's
structure through interactions with the hard walls. Analysis of the
resulting data requires a detailed study of the dependence on $L$.
In this setup, the expectation value of the hadron velocity vanishes
for some directions $i$, $\langle v_i \rangle =0$. Keeping in mind the
caveat already raised in section \ref{v0seca}, that $\langle v \rangle =0$
is a considerably weaker condition than the presence of a true $|v=0\rangle $
or $|v^2 =0\rangle $ state, the resulting hadron wave functions have the
character of anisotropic, hybrid zero-velocity/zero-momentum states.
While the $v=0$ results (\ref{v0result}) are thus not directly applicable,
the developments in section \ref{v0sec} provide a calculational scheme
that can be adapted to treat such more specialized scenaria, together with
an indication of their possible limitations.

More complicated spatial structures are induced for charged particles in
a magnetic background field, which are described by Landau levels in the
plane perpendicular to the magnetic field. Significant effort is required
to achieve a good overlap with these states in lattice calculations,
through an $SU(3)\times U(1)$ eigenmode projection technique~\cite{adel3},
along with a hadronic Landau eigenmode projection to the lowest Landau level,
to extract polarizabilities for charged pions and the proton
\cite{adel4,adel5,adel6}. Due to the localization of the Landau levels
in the directions perpendicular to the magnetic field (whereas propagation
along the direction of the field is free), the charged hadron wave functions
again have the character of hybrid zero-velocity/zero-momentum states,
and the analysis of possible Foldy-type effects would require further,
more specific study that lies beyond the scope of the present work.

In view of the manifold details characterizing any particular lattice
setup, no exhaustive conclusions can be drawn concerning the accounting
for Foldy-type effects in Lattice QCD polarizability calculations.
Nonetheless, we anticipate the two scenaria studied here to be relevant
for the analysis of selected computations, a case in point being further
developments within the zero-momentum scheme developed to study the
dipole \cite{elpol,polprog} and spin polarizabilities \cite{spinpol}.

%%%%%%%%%%%%%%%%%%%%%

\section*{Acknowledgments}
Fruitful discussions with S.~A.~Coon, W.~Detmold, H.~Grie\ss hammer and
B.~Pasquini are acknowledged.
This research was supported by the Erwin Schr\"odinger Fellowship program
of the Austrian Science Fund FWF (``Fonds zur F\"orderung der
wissenschaftlichen Forschung'') under Contract No. J3425-N27 (R.H.)
and the U.S.~Department of Energy, Office of Science, Office of Nuclear
Physics through grant DE-FG02-96ER40965 (M.E.,J.S.).

\end{document}